\documentclass[fleqn,usenatbib]{mnras}
\usepackage{newtxtext,newtxmath}
\usepackage[T1]{fontenc}
\usepackage{ae,aecompl}
\usepackage{graphicx}	
\usepackage{amsmath}	
\usepackage{amssymb}	
\usepackage{color}
\usepackage{longtable}
\usepackage{supertabular,booktabs}
\usepackage[utf8]{inputenc}
\usepackage{url}
\usepackage{multirow}
\usepackage{scrextend}
\usepackage{url}
\usepackage[flushleft]{threeparttable}
\usepackage{supertabular}
\usepackage{braket}

\newcommand{\fev}{{FeV}}
\newcommand{\vp}{{\sc vpfit}}


\title{Constraining the magnetic field on white dwarf surfaces; Zeeman effects and fine structure constant variation}

\author[J.Hu et al]{
J. Hu$^{1}$,
J. K. Webb$^{1}$\thanks{E-mail: jkw@phys.unsw.edu.au},
T. R. Ayres$^2$,
M. B. Bainbridge$^3$,
J. D. Barrow$^4$, \newauthor
M. A. Barstow$^3$,
J. C, Berengut$^1$,
R.F. Carswell$^5$,
V. A. Dzuba$^1$,
V. V. Flambaum$^1$, \newauthor
J. B. Holberg$^{6}$,
C. C. Lee$^4$,
S. P. Preval$^3$,
N. Reindl$^3$,
W.-Ü L. Tchang-Brillet$^7$.\\ \\
$^{1}$School of Physics, University of New South Wales, Sydney, NSW 2052, Australia\\
$^{2}$Center for Astrophysics and Space Astronomy, University of Colorado, 389 UCB, Boulder, Colorado 80309-0389, USA\\
$^{3}$Department of Physics and Astronomy, University of Leicester, University Road, Leicester LEI 7RH, UK\\
$^{4}$DAMTP, Centre for Mathematical Sciences, University of Cambridge, Cambridge CB3 0WA, UK\\
$^{5}$Institute of Astronomy, Madingley Road, Cambridge CB3 0HA, UK\\
$^{6}$Lunar and Planetary Laboratory, Sonett Space Science Building, University of Arizona, Tucson, Arizona 85721, USA\\
$^{7}$LERMA, Observatoire de Paris-Meudon, PSL Research University, CNRS UMR8112, Sorbonne Université, F-92195 Meudon, France
}

\date{Accepted XXX. Received YYY; in original form ZZZ}
\pubyear{2018}

\begin{document}
\label{firstpage}
\pagerange{\pageref{firstpage}--\pageref{lastpage}}
\maketitle

\begin{abstract}
White dwarf atmospheres are subjected to gravitational potentials around $10^5$ times larger than occur on Earth.  They provide a unique environment in which to search for any possible variation in fundamental physics in the presence of strong gravitational fields.  However, a sufficiently strong magnetic field will alter absorption line profiles and introduce additional uncertainties in measurements of the fine structure constant. Estimating the magnetic field strength is thus essential in this context. Here we model the absorption profiles of a large number of atomic transitions in the white dwarf photosphere, including first-order Zeeman effects in the line profiles, varying the magnetic field as a free parameter. We apply the method to a high signal-to-noise, high-resolution, far-ultraviolet HST/STIS spectrum of the white dwarf G191-B2B. The method yields a sensitive upper limit on its magnetic field of $B < 2300$ Gauss at the $3\sigma$ level. Using this upper limit we find that the potential impact of quadratic Zeeman shifts on measurements of the fine structure constant in G191-B2B is 4 orders of magnitude below laboratory wavelength uncertainties.
\end{abstract}

\begin{keywords}
white dwarfs -- magnetic fields -- line: profiles -- atomic processes
\end{keywords}

\section{Introduction}

The motivation for this study arises in the context of searching for any changes in fundamental constants in the photosphere of a white dwarf, an environment where the gravitational potential is $\sim 10^4 - 10^5$ that on Earth \citep{berengut13, bainbridge17}. A range of different theoretical models have been considered in which a cosmological light scalar field representation of the fine structure constant $\alpha$ can couple to gravity, see e.g. \citet{magueijo02}.

Constraining the fine structure constant variation requires accurate measurements of the relative wavelength separations between atomic transitions. Most known white dwarfs ($> 80$\%) are classified as ``non-magnetic'', e.g.  \citep{kawka07,sion14,kepler13,kepler15}. However, the extent to which stars classified as non-magnetic lack any field at all has recently been examined \citep{Scaringi2017}. Detecting the presence of a weak magnetic field, or placing an upper limit is important because the linear Zeeman effect impacts on spectral line shapes and the quadratic Zeeman effect can produce relative line shifts. Both impact on modelling spectral lines and hence contribute to the overall uncertainty on $\alpha$ measurements.

In the large sample of currently known, magnetic white dwarfs exhibit field strengths in the range from a few time $10^3$ Gauss up to $\sim 10^9$ Gauss. At the weak field end of this range, spectropolarimetry measurements have proved most sensitive. A recent extensive survey using spectropolarimetry measurements by \citet{bagnulo18} found no floor below which magnetic fields dissipate and at the sensitivity of their survey were able to place upper limits on non-detections at the 1-2 kG level.

In this paper we illustrate a method for deriving sensitive limits on the magnetic field strength using only spectral lines i.e. without the need for spectropolarimetry data. The method makes simultaneous use of multiple atomic transitions. When the magnetic field is sufficiently weak, Zeeman splitting may be unresolved. The number of Zeeman components and their relative separations depends on the angular momentum quantum numbers of the upper and lower states involved in the transition and these vary from one absorption line to another. This is useful because it means that by simultaneously modelling multiple absorption lines one can disentangle magnetic field effects from other line broadening effects.

In the following sections, we estimate the magnetic field in the photosphere of the white dwarf G191-B2B, using high spectral resolution Hubble Space Telescope observations.

\section{Astronomical Data}
\label{spectrum}

The data used in this analysis were obtained by the Hubble Space Telescope (HST) and its Space Telescope Imaging Spectrograph (STIS), mostly with the high-resolution far-UV echelle grating (E140H: resolving power $\lambda/(2\Delta\lambda) \approx 114,000$) and the FUV-MAMA detector, although one setting of grating E230H with the NUV-MAMA (same resolving power) to extend coverage past 1700{\AA}. The spectra were collected from the Mikulski Archive for Space Telescopes (MAST), and cover the approximate wavelength range 1150{\AA} to 1900{\AA}, with the best signal-to-noise over the interval 1150{\AA} to 1670{\AA}. Altogether, 22 E140H spectra totalling 43.9~ks of exposure, and 5 E230H spectra totalling 11.4~ks exposure, were incorporated in the analysis.  All the observations were through the $0.2\arcsec{\times}0.2\arcsec$ ``photometric'' aperture.

\begin{figure*}
\centering
\includegraphics[width=\textwidth]{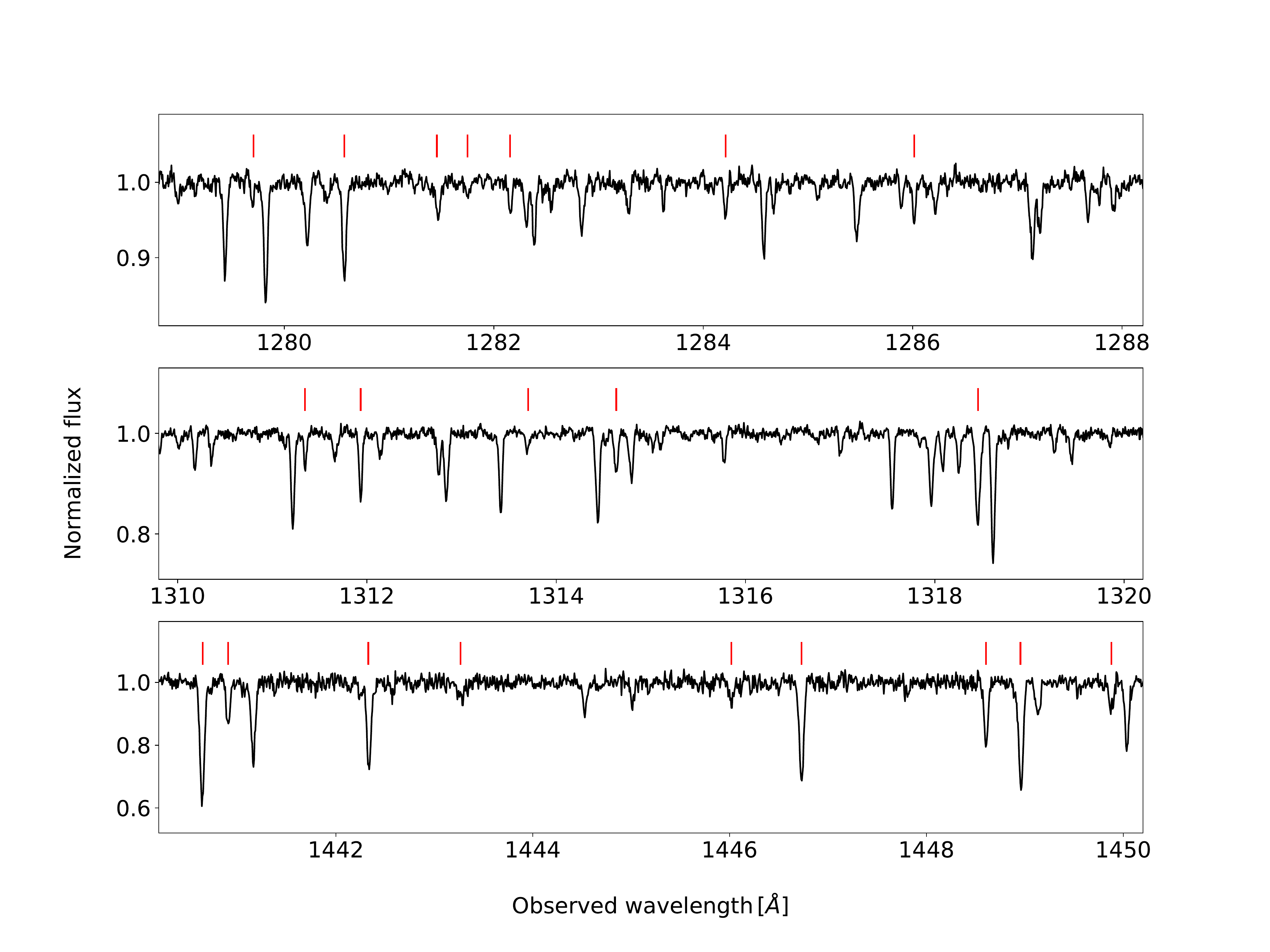}
\caption{Example segments of the HST/STIS high-resolution FUV echelle spectrum of G191-B2B. It was obtained from the co-addition of 27 exposures from the MAST archive, with a total integration time of 55.4~ks.  The spectral resolution (FWHM) is $R \approx 114,000$, and the peak S/N per 2.6 km s$^{-1}$ resel exceeds 100. Red ticks above absorption lines indicate {\fev} transitions used in this analysis. Total wavelength coverage of the G191-B2B UV spectrum is 1150 to 1897 {\AA}, but the absorption line density falls off rapidly beyond about 1730{\AA}.}
\label{fig:segments}
\end{figure*}

The final co-added spectrum of G191-B2B is of very high quality, with a peak signal-to-noise (S/N) per 2.6 km s$^{-1}$ resolution element (``resel'') of better than 100.  A previous analysis of the G191-B2B archival echellegrams, based on an earlier processing and co-addition scheme, was presented in \citet{preval13}. Figure \ref{fig:segments} shows some examples of the {\fev} absorption lines used in the present analysis.

Further details concerning the data reduction, and a table of the archival exposures, can be found in Appendix \ref{AppendixA}.  

\section{Method}\label{method}

The high spectral resolution means that individual absorption lines are resolved.  This facilitates a new method for measuring, or placing an upper limit on, the white dwarf photospheric magnetic field.  The procedure adopted is as follows: 
\begin{enumerate}
\item[1.] Absorption lines are detected above a 5$\sigma$ threshold in the G191-B2B FUV continuum-normalised spectrum;
\item[2.] For each {\fev} line, calculate the first order Zeeman wavelength shifts for a field of strength $B$;
\item[3.] Fit multi-component Voigt profiles\footnote{The {\fev} lines used are optically thin. It was found that profile fitting typically results in normalised $\chi^2$ values around unity so Voigt profiles work well empirically.} (i.e. one Voigt profile for each Zeeman component) simultaneously to all {\fev} lines, using tied parameter constraints as discussed in Section \ref{adata};
\item[4.] Step through a range of $B$, minimising $\chi^2$ for the overall fit to solve for an upper limit on $B$.
\end{enumerate}

\subsection{Zeeman component wavelength calculation}\label{Zeemanlamda}

The wavelengths of the Zeeman components are calculated as follows. The energy shift $\Delta E_{ik}$ from the $B=0$ electric dipole E1 value to the $n^{th}$ first-order Zeeman-split component is characterised by the total angular momentum quantum numbers of the lower and upper state, $J_i$ and $J_k$ and the Landé factors for the lower and upper states, $g_i$ and $g_k$. This is given by
\begin{equation} \label{energy}
\Delta E_{ik}(n) = \mu B \left(g_i m_i - g_k m_k \right)
\end{equation}
where $\mu$ is the electron magnetic moment and the Landé factors $g$ are \citep{aggarwal17},
\begin{equation} \label{lande}
g=1+\sum_{\text{CLS}}{\alpha(\text{CLSJ})\frac{J(J+1)-L(L+1)+S(S+1)}{2J(J+1)}}
\end{equation}
where the summation is over all possible configuration state functions, $C$ is the configuration, $L$ is the orbital angular momentum and $S$ is the spin angular momentum and $J$ is the total angular momentum; $\alpha(\text{CLSJ})$ is a weight (a square of the expansion coefficient) of a particular configuration state function for the level eigen-function.

The $m_i, m_k$ are the projections of the total angular momentum $J_i$ and $J_k$. For each of the lower and upper levels, the permitted number of $m$ states is $(2J+1)$ with $-J \le m \le J$ and selection rule $m_i - m_k = 0, \pm 1$ \citep[Chapter~17]{cowan81}. 

With energy $E$ in cm$^{-1}$ and wavelength in {\AA}, the wavelength $\lambda_n$ of the $n^{th}$ Zeeman component can be calculated as
\begin{equation}\label{new-wav}
\lambda_n = \left( \frac{1}{\lambda_0} + \frac{\Delta E_{ik}(n)}{10^8} \right)^{-1}
\end{equation}

\subsection{Zeeman component oscillator strengths and natural damping constants} \label{Zeemanparams}

To compute model Voigt profiles we need oscillator strengths $f_{ik}$ and natural damping constants $\Gamma_{ik}$ for the $B=0$ E1 line and for the Zeeman-split components. The oscillator strength for a transition between lower and upper levels $i, k$\ is given by \citep{corney77,hilborn02}
\begin{equation} \label{eq:f}
f_{ik} = A_{ki} \frac{w_k}{w_i} \frac{m_e c \epsilon_0 \lambda_{ik}^2}{2 \pi e^2}
\end{equation}
where the statistical weights of the lower(upper) levels are $w_{i(k)} = (2J_{i(k)} + 1)$, $m_e$ is the electron mass, $c$ is the speed of light, $\epsilon_0$ is permittivity of free space, $\lambda_{ik}$ is the transition wavelength, and $e$ is the electron charge.

The Einstein coefficient $A_{ki}$ is the transition probability per second averaged over the upper levels $m_k$ and summed over the lower levels $m_i$
\begin{equation} \label{eq:einstein}
A_{ki} = \frac{1}{w_k}\sum_{m_k}\sum_{m_i}a(J_k,m_k \rightarrow J_i, m_i)
\end{equation}
where $a(J_k,m_k \rightarrow J_i, m_i)$ is the probability per second of radiative transition between two Zeeman sub-levels $m_k$ and $m_i$.

For an electric dipole E1 line \citep[Chapter~14]{cowan81},
\begin{equation} \label{eq:asub}
a(J_k,m_k \rightarrow J_i, m_i) = \frac{16\pi^3}{3h\epsilon_0\lambda^3}|\Bra{J_i,m_i}Q_q^{(1)}\Ket{J_k, m_k}|^2
\end{equation}
where
\begin{equation}
\Bra{J_i,m_i}Q_q^{(1)}\Ket{J_k, m_k} = (-1)^{J_i-m_i}
\left(\begin{matrix}
J_i & 1 & J_k \\
-m_{i} & q & m_{k}
\end{matrix}\right)^{2}
\Bra{J_i}|Q^{(1)}|\Ket{J_k}
\end{equation}
is a matrix element of the electric dipole operator component, the reduced matrix element $\Bra{J_i}|Q^{(1)}|\Ket{J_k}$ is independent of $m_k$ and $m_i$, the  condition $-m_i + q + m_k = 0$ must be satisfied for the Wigner 3-j coefficient to be non-zero, and the 3-j coefficient sum rule results in
\begin{equation} \label{eq:aki}
A_{ki} = \frac{16\pi^3}{3h\epsilon_0\lambda^3w_k}|\Bra{J_i}|Q_q^{(1)}|\Ket{J_k}|^2
\end{equation}
Defining the line strength of the transition $k \rightarrow i$ by $S = |\Bra{J_i}|Q^{(1)}|\Ket{J_k}|^2$
($S_{ki} = S_{ik}$), one finds the relations
\begin{equation}
A_{ki}w_k = \frac{16\pi^3}{3h\epsilon_0\lambda_{ki}^3}S_{ki} \quad\textrm{and}\quad f_{ik} = \frac{S_{ik}}{w_i}\frac{8\pi^2m_ec}{3he^2\lambda_{ik}}
\end{equation}
For the nth pair of Zeeman sub-levels the line strength can be computed as

\begin{equation}\label{3js}
S_{ik(n)} = S_{ik} \left(\begin{matrix} J_i & 1 & J_k \\ -m_{i(n)} & q & m_{k(n)} \end{matrix}\right)^{2}
\end{equation}

The corresponding oscillator strength $f_n$ therefore satisfies
\begin{equation}\label{3jf}
f_{ik(n)}\lambda_{ik(n)} = f_{ik}\lambda_{ik}\left(\begin{matrix} J_i & 1 & J_k \\ -m_{i(n)} & q & m_{k(n)} \end{matrix}\right)^2
\end{equation}

If both upper and lower states are excited states, then the radiative damping constant $\Gamma$ can be expressed as
\begin{equation}\label{gam}
\Gamma_{\text{rad}} = \Gamma_i + \Gamma_k = \sum\limits_{p < i} A_{ip} + \sum \limits_{p < k} A_{kp}
\end{equation}
where $p$ stands for all the possible decay channels of the upper state. According to Equation \ref{eq:aki}, it is clear that the Einstein coefficient is dependent of the transition energy $E_{ki}$, which takes the form
\begin{equation}
A_{ki} \propto \lambda_{ki}^{-3} \propto E_{ki}^3
\end{equation}

If a weak magnetic field $B$ is applied, then
\begin{equation}
\frac{A_{ki}|_{B \ne 0}}{A_{ki}|_{B=0}} = \left( \frac{E_{B \ne 0}}{E_{B=0}} \right)^{3} = \left( \frac{E_{B=0} + \Delta E}{E_{B=0}} \right)^{3} \simeq 1 + 3\left( \frac{\Delta E}{E_{B=0}} \right)
\end{equation}
where the $ki$ subscripts have been dropped for simplicity. For magnetic fields $B \lesssim 50$ kG (see Figure \ref{fig:example_profiles}), where $\Delta v \sim 10$ km/s,
\begin{equation}
\frac{\Delta E}{E_{B=0}} = \frac{\Delta v}{c} \sim 10^{-4},
\end{equation}
implying that it is reasonable to ignore the effect of non-zero $B$ on $A_{ki}$.

We have also only considered a radiative contribution to the damping constant.  However, $\Gamma_{\text{tot}} = \Gamma_{\text{radiative}} + \Gamma_{\text{Stark}} + \Gamma_{\text{collisions}}$. Since the last two terms are likely to be small and in any case including them would only result in a marginally more stringent upper limit on $B$, we have taken them to be zero.

\section{{\fev} atomic data and absorption line profile fitting} \label{adata}

Einstein coefficients $A_{ki}$ for {\fev} are tabulated in \cite{kramida14}. Values for $\Gamma$ are obtained from Kurucz database\footnote{\url{http://kurucz.harvard.edu/atoms.html}}. We use FeV wavelengths from laboratory measurements that have been re-calibrated onto a Ritz scale by \citet{kramida14}, selecting transitions from the online table\footnote{\url{http://iopscience.iop.org/0067-0049/212/1/11/suppdata/apjs493576t1\_mrt.txt}}, which also gives $J$ values.  Values of $g$ for the upper and lower energy levels of lines of interest are taken from Table 1 in \citet{aggarwal17}.

Using the above procedure to define the parameter set for each Zeeman component, {\vp}\footnote{\url{https://www.ast.cam.ac.uk/$\sim$rfc/vpfit.html}} \citep{carswellwebb14} is run multiple times, externally varying the magnetic field $B$.  Each Zeeman component is taken to be a Voigt profile for convenience. Oscillator strengths for individual Zeeman-split components are calculated using equation \ref{3jf}. The damping constant for each Zeeman component approximated as the $B=0$ value, as described above.

Each absorption component is parameterized with column density $N$, redshift $z$ and velocity dispersion parameter $b$, where $b$ is equal to $\sqrt{2}\sigma$, $\sigma$ is the root mean square of the velocity distribution. Here we fit multiple {\fev} lines simultaneously, where $\log N$ each transition (not each Zeeman component) is treated as a free parameter. The redshift $z$ of each transition is also treated as a free parameter so any additional effects that may cause line shifts will be accounted for. 

When modelling the data, we make a simplifying assumption concerning line width that reduces degeneracy between line width $b$ and the magnetic field strength $B$: the $b$-parameters for all {\fev} lines are constrained to be the same. However \cite{barstow99} have shown that different {\fev} transitions arise at different atmospheric depths. Nevertheless, the detected {\fev} transitions form over a tiny range in atmospheric scale-height (a few kilometers) compared to the white dwarf radius ($\sim 14,000$ km). Using the {\sc TLUSTY v200} stellar atmosphere model for G191-B2B described in \cite{preval13}, we checked whether the constant $b$ assumption was reasonable by grouping {\fev} transitions into four stratified layers, tying $b$ to be the same for all lines within each layer, and re-fitting. All four $\langle b \rangle$ values were consistent suggesting the original constant $b$ assumption is reasonable. Therefore, fitting $n$ {\fev} lines, there are a total of $2n+1$ degrees of freedom for our model for each value of the magnetic field $B$.

When fitting absorption profiles to the data, the non-Gaussian HST/STIS instrumental profile was taking into account using numerical convolution.  The instrument profile is accurately measured at 1200{\AA} and 1500{\AA}\footnote{\url{http://www.stsci.edu/hst/stis/performance/spectral_resolution}}. We linearly interpolated from these 2 measured functions to the observed wavelength of each absorption line model.

\subsection{{\fev} lines used}

We defined the sample of {\fev} lines used as follows.  First lines were detected in the HST FUV spectrum $\geq 5\sigma$ significance. The observed wavelengths of those lines were then shifted to their rest-frame values using a redshift $z=0.000079$ using the G191-B2B radial velocity measured by \cite{preval13}. The wavelength of each observed line was then compared to the {\fev} list in \cite{kramida14} and accepted as an identification if the wavelength differences were $\leq 5\sigma$.  Finally, each line was then fitted individually, with $z$, $b$, and $\log N$ as free parameters, and with $B$ allowed to vary in the range $0 < B << 20$ kG.  If, for any value of $B$, the normalised $\chi^2$ for any fit did not fall below 1.1, the line was rejected. This last step aimed to eliminate {\fev} lines blended with other species. These selection criteria resulted in a final sample of 119 {\fev} lines.

\subsection{Example transitions}

\begin{figure*}
\includegraphics[width=\textwidth]{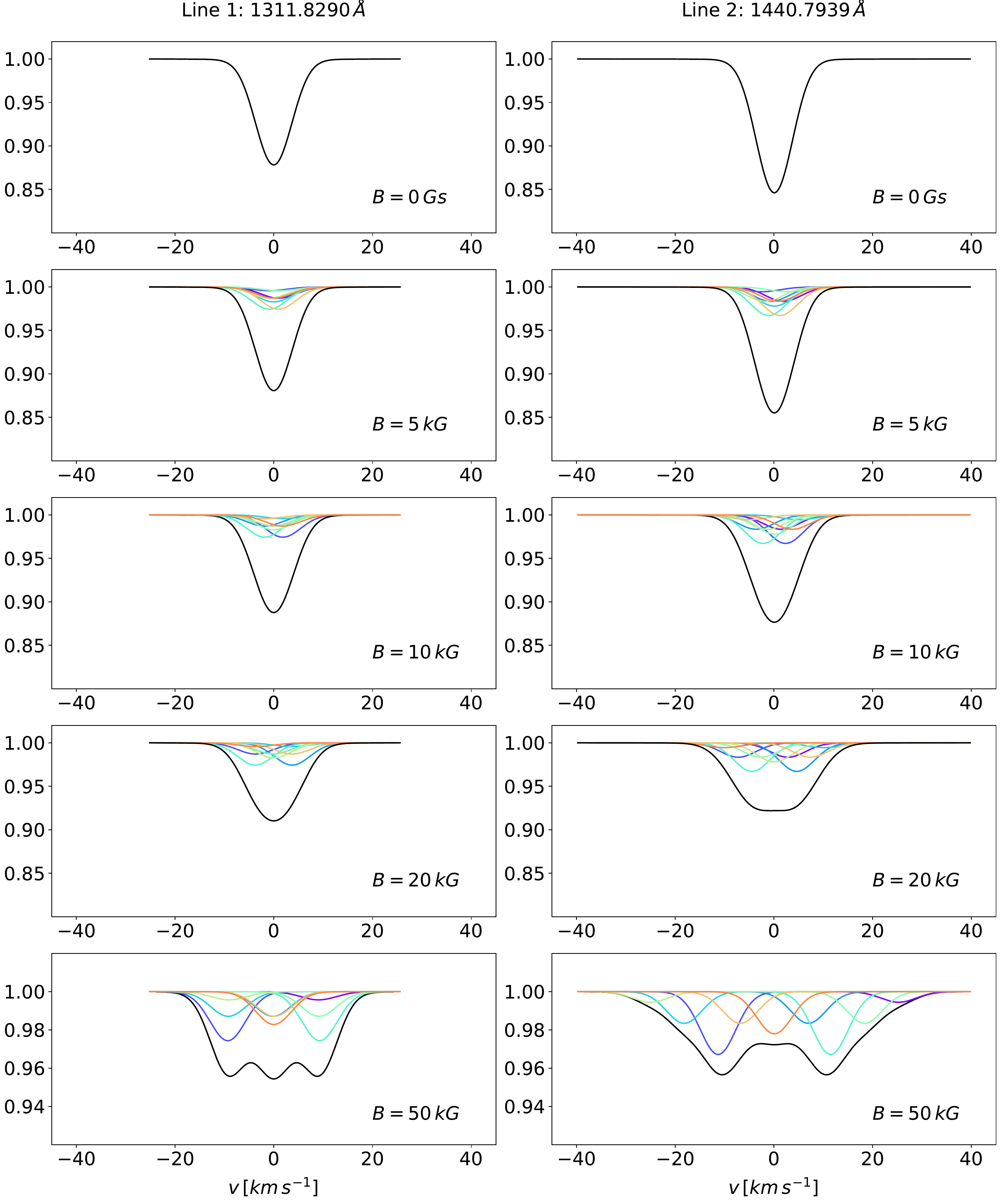}
\caption{{\bf Line profiles for different magnetic field strengths.} The constraint on $B$ is obtained by simultaneously fitting a total of 119 lines. Here we  plot two example profiles that show how the Zeeman structures can vary substantially from one line to another. The left hand column illustrates the {\fev} 1131.8290{\AA} transition and the right hand column  {\fev} 1440.7939{\AA} (rest-frame wavelengths). The line component widths in this illustration are set to $b=5$ km/s, comparable to the measured values.  The line strengths (column densities) are observed to be slightly different for the two lines, hence the different line depths for the two lines. The $y$-axis is normalised flux. The profiles are plotted in velocity units, $v = c(\lambda - \lambda_0)/\lambda_0$, where $\lambda_0$ is the line central wavelength.  The magnetic field strength is increased from zero in the top panel to $B=50$kG in the lowest panel. Table \ref{tab:sample} lists the details for these two example lines.}
\label{fig:example_profiles}
\end{figure*}

We pick two transitions to illustrate the variation in Zeeman structure from one line to another as $B$ increases. The 1311.8290\AA~ line has quantum numbers $J_i=1$, $J_k=2$ and Landé factors $g_i=1.001$, $g_k=0.989$. The 1440.7939\AA~ line has $J_i=1$, $J_k=2$, $g_i=2.494$, $g_k=1.811$. Both lines split into 9 components. Figure \ref{fig:example_profiles} shows these two example profiles for various values of the field strength $B$.  Component structure is only resolved beyond $B=20$kG. It is this differential profile structure that allows a constraint on $B$ to be derived by simultaneously fitting multiple lines.

\begin{table}
\caption{Zeeman first-order splitting for two example {\fev} electric-dipole transitions with wavelengths 1311.8290 and 1440.7939{\AA} for 5 values of the magnetic field strength $B$. Both lines split into 9 components but the 2 profiles become increasingly different as $B$ increases, as Figure \ref{fig:example_profiles} shows.}
\label{tab:sample}
\begin{tabular}{c |c c | c c}
\hline\hline
\textbf{$B\,[kG]$} & $\lambda\,$[{\it \AA}] & \textbf{$f$} & $\lambda\,$[{\it \AA}] & \textbf{$f$} \\ 
\hline \hline
\rule{0pt}{4ex} 
  0 & 1311.8290 & 0.19000 & 1440.7939 & 0.16000\\ [2ex]
  \hline
  \rule{0pt}{4ex}   
  \multirow{9}{*}{5} & 1311.8249     & 0.03800 & 1440.7818     & 0.00533 \\
& 1311.8250     & 0.01900 & 1440.7851     & 0.01600 \\
& 1311.8250     & 0.00633 & 1440.7884     & 0.03200 \\
& 1311.8290     & 0.02533 & 1440.7906     & 0.01600 \\
& 1311.8290     & 0.01900 & 1440.7939     & 0.02133 \\
& 1311.8290     & 0.01900 & 1440.7972     & 0.01600 \\
& 1311.8330     & 0.01900 & 1440.7994     & 0.03200 \\
& 1311.8330     & 0.00633 & 1440.8027     & 0.01600 \\
& 1311.8331     & 0.03800 & 1440.8060     & 0.00533 \\[2ex]
   \hline
  \rule{0pt}{4ex}  
  \multirow{9}{*}{10} &	1311.8209 &	 0.03800 & 1440.7697   &    0.00533 \\
&	1311.8210 &      0.01900 & 1440.7763 &    0.01600\\
&	1311.8211 &      0.00633 & 1440.7830 &    0.03200\\
&	1311.8289 &      0.01900 & 1440.7873 &    0.01600\\
&	1311.8290 &      0.02533 & 1440.7939 &    0.02133\\
&	1311.8291 &      0.01900 & 1440.8005 &    0.01600\\
&	1311.8369 &      0.00633 & 1440.8048 &    0.03200\\
&	1311.8370 &      0.01900 & 1440.8115 &    0.01600\\
&	1311.8371 &      0.03800 & 1440.8181 &    0.00533\\ [2ex]
   \hline
  \rule{0pt}{4ex}
  \multirow{9}{*}{20} &   1311.8127 &	  0.03800 & 1440.7456  &  0.00533  \\
&   1311.8129 &   0.01900 & 1440.7588  &  0.01600  \\
&   1311.8131 &   0.00633 & 1440.7720  &  0.03200  \\
&   1311.8288 &   0.01900 & 1440.7807  &  0.01600  \\
&   1311.8290 &   0.02533 & 1440.7939  &  0.02133  \\
&   1311.8292 &   0.01900 & 1440.8071  &  0.01600  \\
&   1311.8449 &   0.00633 & 1440.8158  &  0.03200  \\
&   1311.8451 &   0.01900 & 1440.8290  &  0.01600  \\
&   1311.8453 &   0.03800 & 1440.8422  &  0.00533  \\ [2ex]
   \hline
  \rule{0pt}{4ex}   
  \multirow{9}{*}{50} &  1311.7883 &   0.03800 & 1440.6731  &  0.00533 \\
&  1311.7888 &   0.01900 & 1440.7061  &  0.01600 \\
&  1311.7893 &   0.00633 & 1440.7392  &  0.03200 \\
&  1311.8285 &   0.01900 & 1440.7608  &  0.01600 \\
&  1311.8290 &   0.02533 & 1440.7939  &  0.02133 \\
&  1311.8295 &   0.01900 & 1440.8270  &  0.01600 \\
&  1311.8687 &   0.00633 & 1440.8486  &  0.03200 \\
&  1311.8692 &   0.01900 & 1440.8817  &  0.01600 \\
&  1311.8697 &   0.03800 & 1440.9148  &  0.00533 \\ [2ex]
\hline\end{tabular}

\end{table}

\section{Results and implications for fine structure constant measurements}
\label{Results}

The motivation for the analysis described in this paper has been to examine the importance of magnetic field effects on measurements of the fine structure constant on white dwarf surfaces. We used the first order Zeeman effect, applied simultaneously to multiple spectroscopic transitions, to constrain $B$. The method provides sensitive upper limits independent of and without the need for separate polarization observations.

\begin{figure*}
\includegraphics[width=\textwidth]{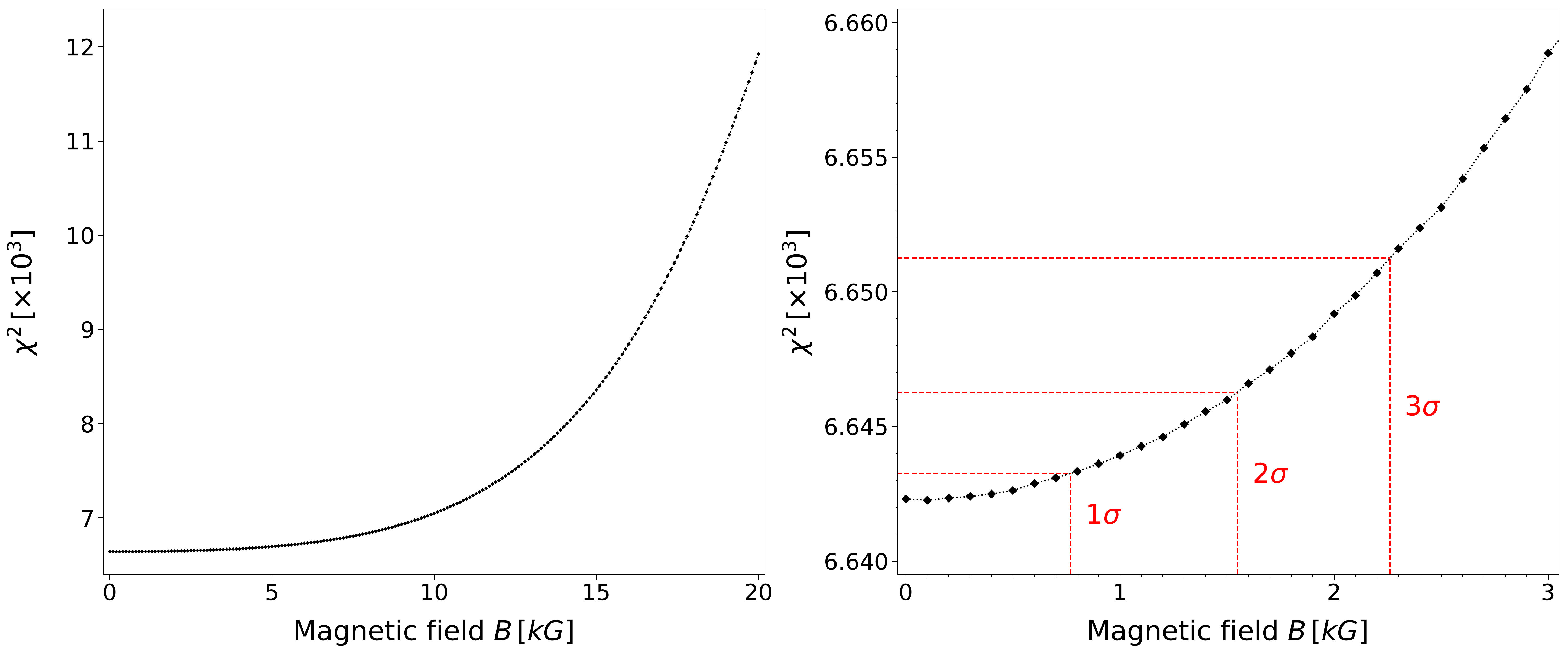}
\caption{Constraining the magnetic field strength $B$ in the photosphere of G191-B2B. The minimum total $\chi^2$ for fit to 119 {\fev} occurs at $B=100$G. Left panel: full $B$ range fitted. Right panel: zoom-in to illustrate 1, 2, and 3$\sigma$ upper limits, given by $(\chi_{min} + \Delta\chi^2)$ where $\Delta\chi^2 = 1, 4, 9$. See Table \ref{tab: bfield-esti-01}.}
\label{fig: magnetic-all}
\end{figure*}

Using a sample of 119 {\fev} absorption lines detected in the HST/STIS FUV spectrum of the white dwarf G191-B2B, we derived an upper limit on $B$, shown in Figure \ref{fig: magnetic-all}, where the minimum $\chi^2$ is seen to occur at $B=100$G.  The method of \citet{lampton76}, using $\chi^2 + \Delta\chi^2$ for one degree of freedom (in our case) gives 1, 2, and 3-$\sigma$ upper limits on $B$ of 800, 1600, and 2300G respectively, as summarised in Table \ref{tab: bfield-esti-01}.  Our result is consistent with a recent previous measurements of G191-B2B by \citet{bagnulo18} who report $B = -280 \pm 965$G.

\begin{table}
\centering
\caption{Final results: 1, 2, and 3$\sigma$ upper limits on the magnetic field $B$ in the G191-B2B photosphere. The model is calculated across the full range $0 < B < 20$ kG in $100$G intervals. See Figure \ref{fig: magnetic-all}.}
\begin{tabular}{c| c| c}
\hline 
\hline 
\multicolumn{1}{c|}{\textbf{Significance Level}} & 
\multicolumn{1}{c|}{\textbf{$\chi^2$}} &
\multicolumn{1}{c}{\textbf{$B$ (G)}} \\
\hline 
1$\sigma$ & 6643.26 & 800 \\
2$\sigma$ & 6646.26 & 1600 \\
3$\sigma$ & 6651.26 & 2300 \\
\hline
\end{tabular}
\label{tab: bfield-esti-01}
\end{table} 

We have focused only on {\fev} in this analysis because many transitions are available and because their laboratory wavelength precision is relatively high. One could in principle include lines from other species (e.g. FeIV and NiV) and improve the constraint on $B$.

We have tested the effect of Stark broadening by modifying the damping constant to include an additional term, $\Gamma = \Gamma_{\text{rad}} + \Gamma_{\text{stark}}$.  $\Gamma_{\text{stark}}$ is obtained using $(\Gamma_{\text{stark}}/n_e)\times n_e$, where $(\Gamma_{\text{stark}}/n_e)$ is from the Kurucz database and the electron density $n_e$ is obtained from the \citet{preval13} model atmosphere. We found that this modifications had a negligible change on any of the results.

We have not taken into account any possible variation in the magnetic field strength across the white dwarf surface. A non-uniformity of this sort would cause Zeeman splitting to vary across the stellar surface in which case the observed line profiles would be formed from the integrated effect across the surface.  If the white dwarf magnetic field is dipolar \citep{kawka18}, the observed line profiles will also depend on the orientation of the magnetic field axis with respect to our vantage point \citep{stibbs50}. However, to some extent this effect may be diluted because the G191-B2B spectrum used in this paper was formed from the co-addition of observations collected over a period of around a decade. If the rotation period is much shorter than 10 years \citep{angel81} and if the stellar rotation axis perpendicular to or aligned with our line of sight, some degree of randomisation of the non-uniform magnetic field effect will be present in the final co-added spectrum.

The quadratic Zeeman effect  is known to be important in white dwarfs with strong magnetic fields  \cite{Angel1974}. The upper limit on $B$ derived for G191-B2B allows us to constrain line shifts associated with quadratic Zeeman effects (as opposed to first order line splitting Zeeman effects). An estimate of the quadratic Zeeman shift  may be obtained using the formula
\begin{equation} \label{quad}
\Delta \lambda_q(\textrm{\AA}) = -4.98 \times 10^{-23} \lambda^2 n_k^4 (1+m_k^2) B^2
\end{equation}
\citep{Jenkins1939,Preston1970,Trimble1971} where $n_k$ is the principal quantum number, $m_k$ is the principal magnetic quantum number (both for the upper level), and $B$ is in Gauss.

Taking an illustrative wavelength of 1300{\AA}, $n_k = 4$\footnote{\url{http://www.pa.uky.edu/~peter/newpage/}}, $m_k = 1$, and $B=2300$ Gauss, $\Delta\lambda_q = 2.3 \times 10^{-4}$m{\AA}. This is $\sim 4$ orders of magnitude smaller than the best available FeV laboratory wavelength measurement uncertainty. In fact, since this calculation used the 3-$\sigma$ upper limit on $B$, the actual shifts are probably even smaller. We can thus safely conclude that FeV quadratic Zeeman shifts are negligible in measurements of the fine structure constant for magnetic fields as small as in G191-B2B. Note that there is a small second order contribution to the quadratic Zeeman shift due to mixing of different excited states by the magnetic field. The magnitude of this paramagnetic shift is comparable to the diamagnetic shift (Equation \ref{quad}) and does not change our conclusion.

\bibliographystyle{mnras}
\bibliography{paper2}

\section*{Acknowledgements}
This research project was undertaken with the assistance of resources and services from the National Computational Infrastructure (NCI), which is supported by the Australian Government.

Based on observations made with the NASA/ESA Hubble Space Telescope, obtained from the data archive at the Space Telescope Science Institute. STScI is operated by the Association of Universities for Research in Astronomy, Inc. under NASA contract NAS 5-26555. 

JH is grateful for a CSC Scholarship and a UNSW tuition fee scholarship. JDB thanks the STFC for support. CCL thanks the Royal Society for a Newton International Fellowship. JKW thanks the John Templeton Foundation, the Department of Applied Mathematics and Theoretical Physics and the Institute of Astronomy Cambridge for hospitality and support, and Clare Hall Cambridge for a Visiting Fellowship.  TRA thanks the STScI for its support, especially through the ASTRAL project. NR is supported by a Royal Commission 1851 research fellowship. WULTB wishes to acknowledge supports from the French CNRS-PNPS national program and from the LabEx Plas@Par managed by the French ANR (ANR-11-IDEX-0004-02). SPP, MAB, and MBB acknowledge the support of the Leverhulme Trust.

\appendix
\section{G191-B2B data reduction details}
\label{AppendixA}

\subsection{Archival STIS echelle data}

G191-B2B is a calibration target for STIS, specifically for the echelle blaze function profiles.  There have been two major echelle campaigns to observe the WD, one at the beginning of the HST mission (1998--2002, mostly 2000 February/March and 2001 September), and a second (2009 November and 2010 January) shortly after Servicing Mission 4, which restored STIS to operational condition after a 5-year hibernation following an electrical failure in 2004. \\

In each of these calibration campaigns, the WD was observed in essentially all 44 supported echelle settings, split between medium and high- ($\lambda/(2\Delta\lambda) \approx 30000 - 45000$, and 114,000, respectively), and the two camera-specific wavelength bands, FUV (1150--1720 \AA) and NUV (1650--3200 \AA).  There were about 150 individual exposures.  Of these, 39 were taken in various FUV high- (E140H) settings. Two of those exposures were very short, however, and not considered further, leaving 37 datasets in FUV/H.  Exposure times ranged from a few hundred seconds up to about 6~ks, with an average of about 1700~s.  An additional five exposures, taken with the NUV high- echelle E230H in its shortest wavelength setting (1763{\AA}), were included to extend coverage past 1700{\AA} (the spectral density of absorption features in G191-B2B falls precipitously beyond about 1730{\AA}), and improve S/N in the 1630--1670{\AA} region that overlaps with FUV settings 1562{\AA} and 1598{\AA}.   \\

The final collection of archival E140H (and E230H) exposures of G191B2B was download from the MAST archive in the form of the raw science images and associated wavelength calibrations (short exposures of an on-board emission-line lamp).  Each raw plus wavecal pair initially was passed through a modified version of the CALSTIS pipeline to yield the so-called x1d file, a tabulation of fluxes and photometric errors versus wavelength for each of the up to several dozen echelle orders of a given grating setting. The modifications consisted of upgrades to several key reference files that control the spectral extractions and calibrations. For example, the existing dispersion relation file was replaced with a new version based on a re-analysis of wavelength calibration material, including deep exposures of the wavecal lamps taken for that purpose; all specialized to the post-SM4 formats of the various echelle settings (the Mode Select Mechanism ``home positions'' have been fixed since 2009, contrary to the earlier operational period [1997--2004] when they periodically were shifted by small amounts to distribute light levels more evenly over the detectors, to counter aging effects).  There also were updates to the locations of the echelle orders for each wavelength setting, based on measurements of post-SM4 G191-B2B raw images. Some of the setting-dependent files were modified to include orders that appear on the detector in contemporary observations, but which were not included in the default reference files.

The CALSTIS x1d files then were subjected to a number of post-processing steps to: (1) correct the wavelengths for additional small residuals due to the relatively low order of the CALSTIS polynomial dispersion relations; (2) adjust the echelle sensitivity (“blaze”) function empirically based on achieving the best match between fluxes in the overlap zones between adjacent orders, as averaged over all the orders with significant S/N; and (3) merge the overlapping portions of adjacent echelle orders to achieve a coherent 1-D spectral tracing for that specific setting. During this step, it was noticed that a number of the echellegrams, exclusively from the initial calibration period 1998--2001, suffered from poor blaze corrections or other defects, likely attributable to the fact that the new CALSTIS reference files were specialized to results from the circa 2009 calibration campaign.  Out of the 28 exposures from the earlier period, 17 were eliminated from further consideration (including the two very short exposures previously discarded).  The loss in S/N was relatively minor, however, because several of the eliminated exposures were relatively short compared with those from the second calibration campaign.  (The retained datasets are marked by star symbols in Table \ref{tab:spectra}.)\\

Thereupon followed a series of steps to merge the up to several independent exposures in a given setting, and finally splice all the setting-specific tracings into a full-coverage high- spectrum.  A general description of the co-addition/splicing protocols can be found in: \url{http://casa.colorado.edu/~ayres/ASTRAL/} (STIS Advanced Spectral Library [ASTRAL] project).  A fundamental component of the ASTRAL protocols is a bootstrapping approach to provide a precise relative, and hopefully also accurate absolute, wavelength scale; and similarly for the spectral energy distribution.  For a typical stellar target, the bootstrapping would average over only a handful of CENWAVEs, roughly 3 -- 9 over the full FUV + NUV range (lower limit is minimum for medium-resolution; upper limit is minimum for high-resolution).  \\

However, G191-B2B represents an optimum case because the maximum number of settings were available in the FUV, namely the 11 supported high-resolution CENWAVEs, plus the additional single NUV CENWAVE.  The large number of settings and their overlaps tends to randomize any residual errors in the wavelength scales.  \\

The resulting H spectra have uniformly high signal-to-noise: S/N $\approx$ 100, or greater, per resolution element, especially for 1150--1430{\AA}.  There are a few places at the shorter wavelengths where there are unflagged sensitivity blemishes on the FUV MAMA detector, which appear as broad ``absorptions'' in the order-merged spectra.  Because the blemishes appear at different wavelengths in the co-added H and M resolution spectra, they were easy to recognize in a cross-comparison of the independent tracings. \\

\begin{table}
\caption{STIS FUV Echellegrams of G191-B2B}
\label{tab:spectra}
\begin{threeparttable}
\begin{tabular}{rcr}
\hline\hline
\multicolumn{1}{c}{\textbf{Dataset}} & 
\multicolumn{1}{c}{\textbf{UT Start Time}} & \multicolumn{1}{c}{\textbf{$t_{\rm exp}$}} \\ 
& \multicolumn{1}{c}{\textbf{(yyyy-mm-dd.dd)}} & 
\multicolumn{1}{c}{\textbf{s}} \\
\hline \hline
\multicolumn{3}{c}{\rule{0pt}{3ex}1234{\AA}~~(1141--1335{\AA})~~20335~s~~135}\\[1ex]
\midrule
$\star$O57U01030 & 1998-12-17.40  & 2789   \\
$\star$O5I010010 & 2000-03-16.98  & 2279   \\
$\star$O5I010020 & 2000-03-17.04  & 3000   \\
$\star$O5I010030 & 2000-03-17.10  & 3000   \\
O6HB10010 & 2001-09-17.58  &  867   \\
O6HB10020 & 2001-09-17.59  &  867   \\
$\star$OBB001040 & 2009-11-30.36  &  867   \\
$\star$OBB005010 & 2009-12-01.22  & 2200   \\
$\star$OBB005020 & 2009-12-01.28  & 6200   \\
\midrule
\multicolumn{3}{c}{\rule{0pt}{3ex}1271{\AA}~~(1160--1356{\AA})~~696~s~~~27} \\[1ex]
\midrule
O6HB10030 & 2001-09-17.60  &  ~~56   \\
O6HB10040 & 2001-09-17.62  & ~640   \\
$\star$OBB001010 & 2009-11-30.29  & ~696   \\
\midrule
\multicolumn{3}{c}{\rule{0pt}{3ex}1307{\AA}~~(1199--1397{\AA})~~1308~s~~~40}\\[1ex]
\midrule
$\star$O6HB10050 & 2001-09-17.63 &  ~654   \\
$\star$OBB001090 & 2009-11-30.51  & ~654   \\
\midrule
\multicolumn{3}{c}{\rule{0pt}{3ex}1343{\AA}~~(1242--1440{\AA})~~6886~s~~~92}\\[1ex]
\midrule
O6HB10060 & 2001-09-17.64 &  ~686   \\
$\star$OBB001060 & 2009-11-30.41  & ~686   \\
$\star$OBB005030 & 2009-12-01.41 & 3100   \\
$\star$OBB005040 & 2009-12-01.48 & 3100   \\
\midrule
\multicolumn{3}{c}{\rule{0pt}{3ex}1380{\AA}~~(1280--1475{\AA})~~752~s~~~29}\\[1ex]
\midrule
O6HB10070 & 2001-09-17.66 &  ~~33   \\
O6HB10080 & 2001-09-17.69 &  ~719   \\
$\star$OBB001030 & 2009-11-30.34 &  ~752   \\
\midrule
\multicolumn{3}{c}{\rule{0pt}{3ex}1416{\AA}~~(1316--1517{\AA})~~2891~s~~~55}\\[1ex]
\midrule
$\star$O57U01020 & 1998-12-17.35 & 2040   \\
O6HB10090 & 2001-09-17.70 &  ~851   \\
$\star$OBB001050 & 2009-11-30.37 &  ~851   \\
\midrule
\multicolumn{3}{c}{\rule{0pt}{3ex}1453{\AA}~~(1359--1551{\AA})~~1038~s~~~30}\\[1ex]
\midrule
O6HB100A0 & 2001-09-17.71 &  ~809   \\
O6HB100B0 & 2001-09-17.75 &  ~229   \\
$\star$OBB001020 & 2009-11-30.30 & 1038   \\
\midrule
\multicolumn{3}{c}{\rule{0pt}{3ex}1489{\AA}~~(1390--1586{\AA})~~1200~s~~~29}\\[1ex]
\midrule
O6HB100C0 & 2001-09-17.76 & 1263   \\
$\star$OBB0010A0 & 2009-11-30.54 & 1200   \\
\midrule
\end{tabular}

\begin{tablenotes}
     \small
\item Datasets marked with ``$\star$'' were included in the final spectrum.  Cut-in headings list, respectively, central wavelength ($\lambda_{\rm C}$) of each E140H (or E230H) setting, wavelength grasp (in parentheses), total exposure time of the included spectra, and average signal-to-noise per resolution element.  The 0.2\arcsec${\times}$0.2\arcsec\ ``photometric'' aperture was used in all cases. 
\end{tablenotes}
\end{threeparttable}
\end{table}

\begin{table}
\contcaption{STIS FUV Echellegrams of G191-B2B}
\begin{tabular}{rcr}
\hline\hline
\multicolumn{1}{c}{\textbf{Dataset}} & 
\multicolumn{1}{c}{\textbf{UT Start Time}} & \multicolumn{1}{c}{\textbf{$t_{\rm exp}$}} \\ 
& \multicolumn{1}{c}{\textbf{(yyyy-mm-dd.dd)}} & 
\multicolumn{1}{c}{\textbf{s}} \\
\hline \hline
\multicolumn{3}{c}{\rule{0pt}{3ex}1526{\AA}~~(1423--1622{\AA})~~2100~s~~~34}\\[1ex]
\midrule
O6HB100D0 & 2001-09-17.78 &  ~887   \\
O6HB100E0 & 2001-09-17.82 &  ~749   \\
$\star$OBB001070 & 2009-11-30.42 & 2100   \\
\midrule
\multicolumn{3}{c}{\rule{0pt}{3ex}1562{\AA}~~(1463--1661{\AA})~~2134~s~~~30}\\[1ex]
\midrule
O6HB100F0 & 2001-09-17.83 & 1996   \\
$\star$OBB001080 & 2009-11-30.48 & 2134   \\
\midrule
\multicolumn{3}{c}{\rule{0pt}{3ex}1598{\AA}~~(1494--1687{\AA})~~4600~s~~~41}\\[1ex]
\midrule
O57U01040 & 1998-12-17.47 & 2703   \\
O5I011010 & 2000-03-17.18 & 2284   \\
O5I011020 & 2000-03-17.24 & 3000   \\
$\star$O5I011030 & 2000-03-17.30 & 3000   \\
$\star$OBB0010B0 & 2009-11-30.56 & 1600   \\
\midrule
\multicolumn{3}{c}{\rule{0pt}{3ex}1763{\AA}~~(1629--1897{\AA})~~11418~s~~~48}\\[1ex]
\midrule
$\star$O5I013010 &  2000-03-19.13  & 2304 \\
$\star$O5I013020  & 2000-03-19.18  & 3000 \\
$\star$O5I013030 &  2000-03-19.25  & 3000 \\
$\star$O6HB20010 &  2001-09-18.65 &  1314 \\
$\star$OBB002090  & 2009-11-28.55  & 1800 \\
\midrule
\end{tabular}
\end{table}

\subsection{Continuum fitting}
In order to measure or fit absorption line profiles, we need an unabsorbed continuum level for the G191-B2B spectrum. The following procedure for fitting the continuum was used:
\begin{enumerate}
\item[1] The {\sc iraf} task {\sc continuum}, part of the {\sc noao.onedspec} package, was used to manually select regions of the spectrum that visually appeared to be free of any absorption features. Since the number density of absorption lines is quite high, this procedure was time consuming and required a large number of region selections, but ended up producing a better estimate of the continuum.
\item[2] The total length in pixels of the G191-B2B E140H spectrum used in this analysis is 115,815.  In practice we found the continuum fitting worked best by dividing the spectrum into 14 separate regions and treating these regions independently from the point of view of continuum fitting (which also solved a character limitation requirement\footnote{\url{http://stsdas.stsci.edu/cgi-bin/gethelp.cgi?icfit.hlp}}).
\item[3] Cubic splines were then fitted to each of the 14 spectral regions, after which the 14 spectral regions were spliced back together, averaging the fitted continua in the overlap regions. Care was taken to ensure that the edges of overlap regions lay in continuum regions and never within absorption line profiles.
\end{enumerate}

\subsection{Error array correction}
Since we are using $\chi^2$ as a goodness-of-fit measure when modelling the absorption profiles, it is important that the G191-B2B spectral error array is reliable. To check this, we looked at the noise characteristics for the previously selected continuum regions, combining all continuum-only pixels into a continuous array and then splitting that into equal sized segments. We then calculated the mean value of the error array and the standard deviation for segments of 2000 pixels. We found the pipeline error array to be consistent with the observed standard deviation at lower wavelengths but too large for $\lambda$ > 1500\AA. To correct this, the following procedures were carried out:
\begin{itemize}
\item[1] Divide the standard deviation for each segment by the mean value of error array within each segment;
\item[2] Fit this ratio with a cubic spline function interpolated onto the full pixel array, i.e. the ratio in [1] is first put back into it's original wavelength array of 115,815 pixels (leaving gaps corresponding to non-continuum regions); 
\item[3] Multiply the original pipeline error array by this interpolated function.
\end{itemize}

\section{Velocity dispersion parameter distribution} \label{AppendixB}

It became noteworthy during the analysis described in this paper that the spread in the $b$-parameter distribution, peaked at $5.0$ km/s, was somewhat broader than expected on the basis of measurement error alone, i.e. the Fev line widths have some intrinsic scatter.  Figure \ref{fig: b-distribution} illustrates the histogram of $b$-parameters for the 119 FeV lines fitted independently.  Two histograms are shown: the whole distribution and the distribution for a subset of the lines for which measurement errors lie in the range $0.15 < \sigma(b) < 0.45$ km/s. The smaller of the two histograms illustrates that the spread in $b$-parameters is $\sim 2$ km/s yet the measurement errors for that subset is much smaller.

\begin{figure}
\includegraphics[width=0.45\textwidth]{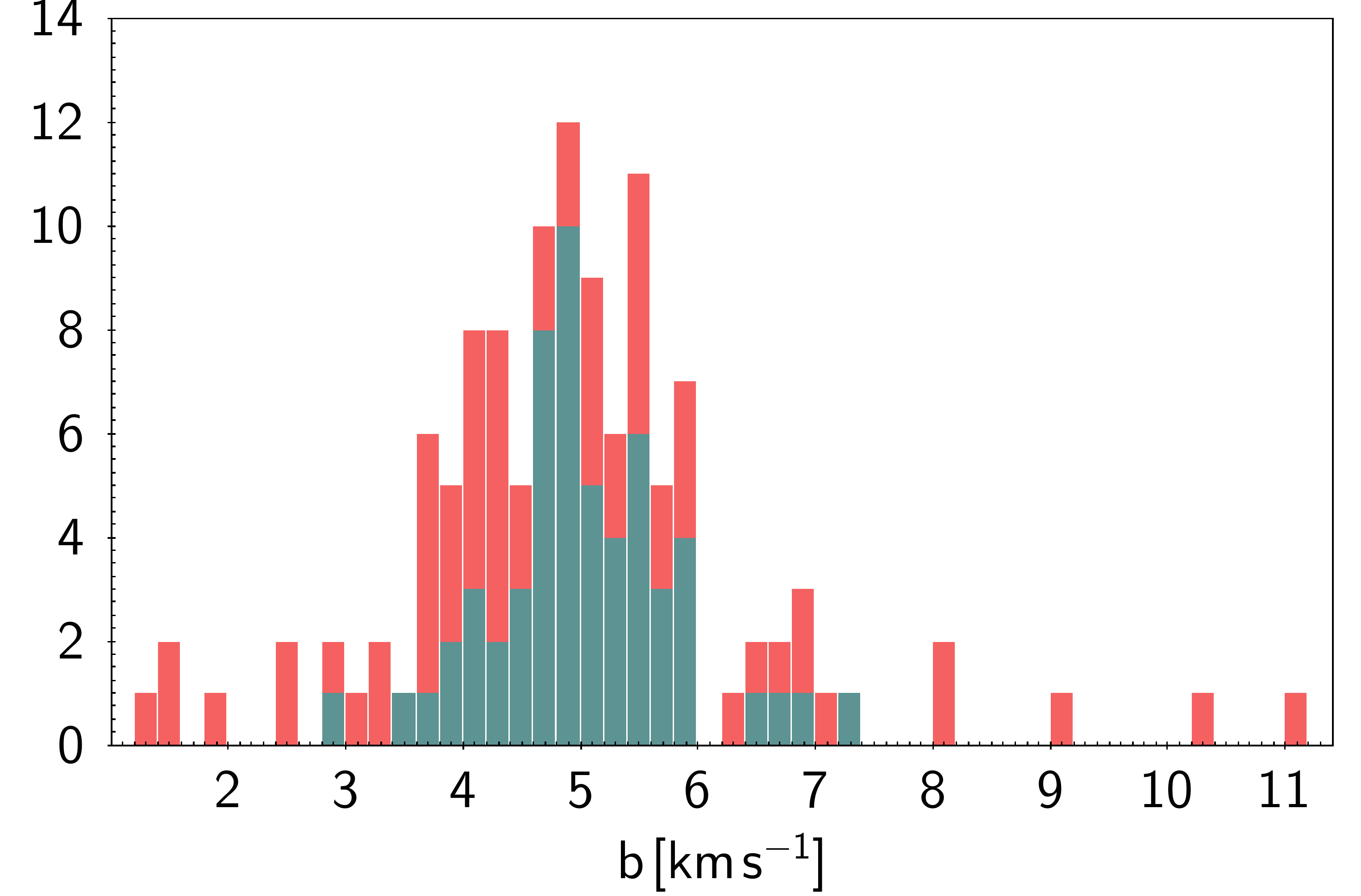}
\caption{The distribution of velocity dispersion $b$-parameters of all 119 FeV lines when fitting individually.}
\label{fig: b-distribution}
\end{figure}

\begin{figure}
\includegraphics[width=0.45\textwidth]{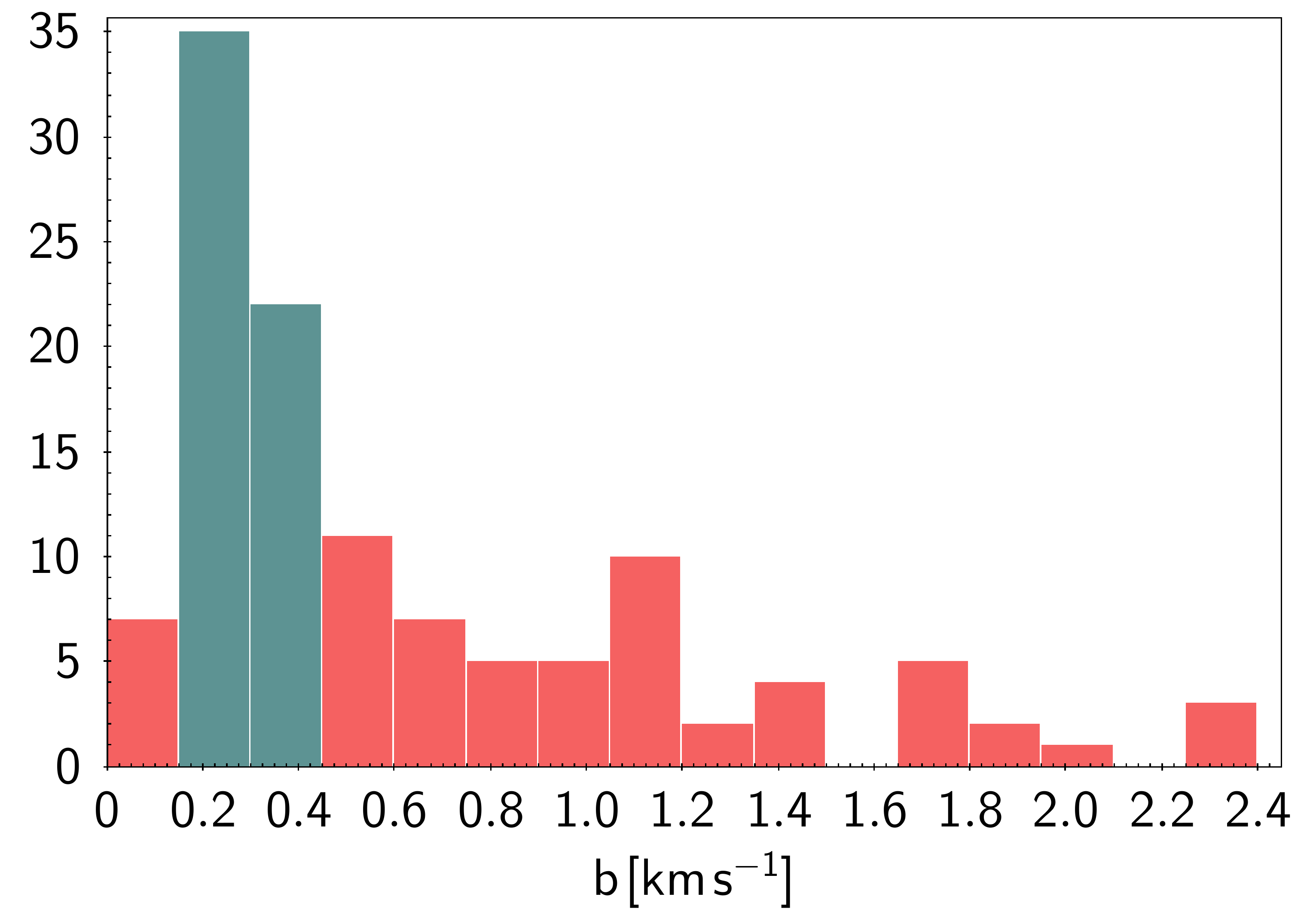}
\caption{The distribution of $b$-parameter measurement errors.}
\label{fig: b-distribution}
\end{figure}

\end{document}